\documentclass[twocolumn,prl,amsmath,amssymb,superscriptaddress,floatfix]{revtex4}
\usepackage[latin1]{inputenc}
\usepackage[english]{babel}
\usepackage[T1]{fontenc}
\usepackage[dvips]{graphicx}
\usepackage{amsmath}
\usepackage{amsfonts}
\usepackage{amssymb}
\usepackage{dcolumn}

\begin{document}
\title{Phase transition to bundles of flexible supramolecular polymers}
\author{B.A.H. Huisman}
\affiliation{Van 't Hoff Institute for Molecular Sciences, University of Amsterdam, Nieuwe Achtergracht 166, 1018 WV Amsterdam, The Netherlands}
\author{P.G. Bolhuis}
\affiliation{Van 't Hoff Institute for Molecular Sciences, University of Amsterdam, Nieuwe Achtergracht 166, 1018 WV Amsterdam, The Netherlands}
\author{A. Fasolino}
\affiliation{Van 't Hoff Institute for Molecular Sciences, University of Amsterdam, Nieuwe Achtergracht 166, 1018 WV Amsterdam, The Netherlands}
\date{\today}
\begin{abstract}
We report Monte Carlo simulations of the self-assembly of supramolecular polymers based on a model of patchy particles. We find a first-order phase transition, characterized by hysteresis and nucleation, toward a solid bundle of polymers, of length much greater than the average gas phase length. We argue that the bundling transition is the supramolecular equivalent of the sublimation transition, that results from a weak chain-chain interaction. We provide a qualitative equation of state that gives physical insight beyond the specific values of the parameters used in our simulations.
\end{abstract}

\maketitle 

Self-assembly is an active field of research, driven by the desire to design
new materials. Understanding the rules of self-assembly has been defined as the challenge for chemistry of this century\cite{Science}. The large molecules involved in self-assembly spontaneously organize because of highly specific interactions like hydrophobic/hydrophilic, hydrogen bonding, Coulomb interaction and $\pi-\pi$-stacking. Modeling these self-assembly processes using coarse-grained models, such as the patchy particle\cite{sciortino}, has the potential to yield theoretical insight. Up to now, patchy-particle models have mostly been used to describe the self-assembly of functionalized colloids for photonic crystals\cite{Wilber}, and to study the formation of self-assembled clusters\cite{Zhang}.

Here, we focus on linear supramolecular self-organi\-zation, such as the reversible aggregation of aromatic molecules and discotic proteins into supramolecular polymers, which has been the subject of several experimental studies\cite{Engelkamp,jonkheijm,Brunsveld}. In this context, Sciortino et al.\cite{sciortino} have shown that the first-order Wertheim Thermodynamic Perturbation Theory (WTPT) of associating liquids accurately predicts the chain-length distribution by comparing the WTPT to simulations of simple square-well patchy particles. One step beyond linear polymerization is the association of supramolecular chains into bundles \cite{Engelkamp,jonkheijm,Brunsveld}, a problem that neither the WTPT, nor the square-well patchy particle, can deal with, since it requires chain-chain interactions. 

In this Letter we present Monte Carlo simulations of a type of patchy particle that, by decreasing temperature or increasing density, first polymerizes into chains and subsequently undergoes a phase transition toward bundles of these chains. We interpret this bundling as a sublimation transition from a polymer gas to a solid bundle. This sublimation competes with polymerization and gives rise to non-trivial phase behavior. We propose a simple thermodynamic model to describe the transition. 

\begin{figure}[t!] 
\begin{center} 
\includegraphics[width=\columnwidth,keepaspectratio]{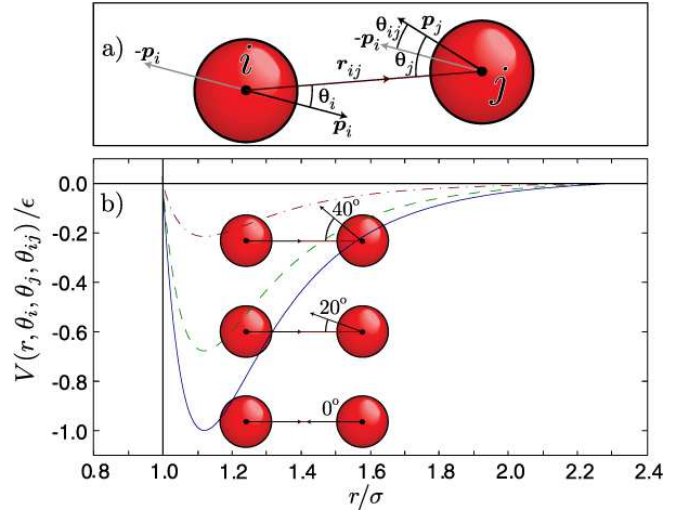}
\end{center} \caption{a) Geometry of the patches. $\theta_i$ is the angle between the direction $\vec{p}_i$ of patch $i$ and the interparticle vector $\vec{r}_{ij}=\vec{r}_{j}-\vec{r}_{i}$. $\theta_j$ is the angle between the direction $\vec{p}_j$ of patch $j$ and $\vec{r_{ji}}=-\vec{r_{ij}}$. $\theta_{ij}$ is the angle between $\vec{p}_j$ and $-\vec{p}_i$. b) Patch potential for three values of $\theta_{j}$ for $w=0.4$. Notice that the minimum at $-1$ occurs for $\theta_i=\theta_j=\theta_{ij}=0$, and that it decreases rapidly for increasing angles. The particles are depicted at half their size for clarity. \label{fig:fig1}}\end{figure}

Our model is a coarse-grained representation of a disc-like, aromatic molecule and consists of hard spheres of diameter $\sigma$, dressed by two opposing patches. The orientationally-dependent patch potential allows not only for chain formation, but also exhibits a weak chain-chain interaction. The patch potential between patch $i$ directed along the vector $\vec{p}_i$ and patch $j$ with direction vector $\vec{p}_j$, illustrated in Fig.~\ref{fig:fig1}, is given by a Lennard-Jones potential of the interparticle distance $r=|\vec{r}_{ij}|$ modulated by three directional components
\begin{equation}
V(r,\theta_i,\theta_j,\theta_{ij})=4\epsilon\left[\left(\frac{\sigma}{r}\right)^{12}-\left(\frac{\sigma}{r}\right)^{6}\right]\exp\left[-\frac{\theta_i^2+\theta_j^2+\theta_{ij}^2}{4w^2}\right]\label{eq:patchLJgauss}
\end{equation}
where $\epsilon$ is the maximum energy of interaction, and $w$ penalizes non-perfect alignment. We truncate $V(r)$ of Eq.~\ref{eq:patchLJgauss} at $r_c=2.3\sigma$ and shift and rescale it to have $V(r_c)=0$ and $V(2^{1/6}\sigma)=-\epsilon$. The first two directional components favor minimization of the angle $\theta_{i}$ between the patch direction $\vec{p_i}$ of patch $i$ and the interparticle vector $\vec{r}_{ij}$, and of the angle $\theta_{j}$ between $\vec{p_j}$ and $\vec{r}_{ji}=-\vec{r}_{ij}$. The third component minimizes the angle $\theta_{ij}$ between $\vec{p}_j$ and $-\vec{p}_i$, favoring parallel alignment of the patches. This potential has three advantages compared to a square-well patch potential\cite{sciortino}. First, it allows multiple bonds without increasing the energy per interaction site in discrete steps. This feature introduces a slight interaction between molecules in neighboring chains. Second, the desired parallel alignment of neighboring patches prevents branching (and therefore network formation) of the polymers. Finally, it allows us to tune the flexibility of a supramolecular chain. One can show that $w$ is a measure of the chain flexibility, by calculating the bending rigidity $\kappa$, and hence the persistence length $l_p$ of a chain \cite{lowe03}
\begin{equation*} 
E_{\text{bend}}=\frac{\kappa}{2}\int_0^L\!\! \frac{1}{R(s)^2} ds\,\Rightarrow\,\kappa=2^{\frac{1}{6}} \frac{3\sigma \epsilon}{4 w^2},\text{ and } l_p\equiv\beta \kappa
\end{equation*}
where $\beta=(k_B T)^{-1}$, $L$ is the chain length, and $R(s)$ is the chain radius at $s$.  For example, the oligo(p-phenylenevinylene)-derivative OPV-4 in dodecane has $l_p= 150\text{nm}$ at $300\text{K}$, which, for a molecule separation of $\sigma=0.35 \text{nm}$ and a bonding energy of $56\text{kJ}/\text{mol}$\cite{jonkheijm} yields $w\approx 0.2$.

We study the thermodynamic equilibrium of a system of $M$ patchy particles in a simulation box of volume $V$ with periodic boundary conditions at temperature $T$ by Monte Carlo simulations. The system is equilibrated by performing moves and rotations of single molecules, and of whole chains. In addition, we perform reptation moves. We note that we can simulate a limited temperature range ($k_B T\gtrsim 0.04\epsilon$) as at lower temperatures the probability of removing a molecule from a chain by a simple Monte Carlo move vanishes. We choose to consider molecules bonded if they interact with an energy $V<-0.3\epsilon$. Contrary to the square-well patch potential \cite{sciortino}, the potential of Eq.\ref{eq:patchLJgauss} makes the average energy per bond $E_{\text{bond}}$ temperature dependent. We find that equipartition (i.e. $E_{\text{bond}}=-\epsilon+3 k_BT$) applies up to $k_B T\lesssim 0.07\epsilon$ where $E_{\text{bond}}\sim -0.79\epsilon$.  

\begin{figure}[t!] 
\begin{center} 
\includegraphics[width=\columnwidth,keepaspectratio]{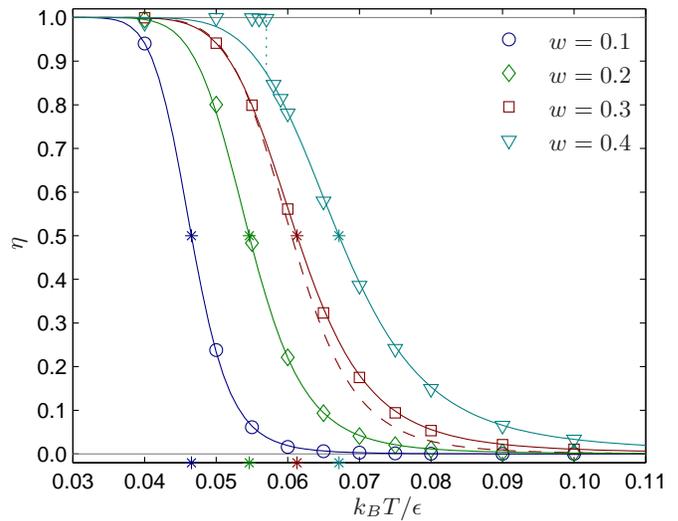}
\end{center} \caption{The aggregation fraction $\eta$ (Eq.~\ref{eq:barNeta}) as a function of temperature for several values of the flexibility $w$ for $M=1330$, $\rho=1.13\cdot 10^{-3}$. $\eta=1/2$ defines the polymerization temperature $T^*$, indicated by an asterisk on the corresponding curves and on the horizontal axis. At $w=0.4$ a transition from a gas of chains, to a solid bundle at $k_B T=0.057\epsilon$ is indicated by the dotted line. The symbols denote simulation results, the solid line is the WTPT. For comparison the dashed line shows a fit of the IFA-model to the curve of $w=0.3$, with $v=1.4\cdot 10^{-4}\sigma^3$ and $G=0.91\epsilon$.  \label{fig:fig2}}\end{figure}

Supramolecular polymerization is reversible and there exists an equilibrium density $\rho(N)$ of chains with length $N$, with $N=1,2,\ldots,\infty$. The total density is given by $\rho=\sum_{N=1}^{\infty} N\rho(N)$. The average chain length $\bar{N}$, and the aggregation fraction $\eta$ are defined as \cite{Ciferri}
\begin{equation}
\bar{N}\equiv\frac{\sum_{N=1}^{\infty} N\rho(N)}{\sum_{N=1}^{\infty} \rho(N) },\,\eta\equiv \frac{\sum_{N=2}^{\infty}N\rho(N)}{\sum_{N=1}^{\infty}N\rho(N)}=1-\frac{1}{\bar{N}^2}\label{eq:barNeta}.
\end{equation}
The fraction of sites that are not bonded $X$, is given by\cite{sciortino}
\begin{equation}
 X=\frac{\sum_{N=1}^{\infty}2\rho(N)}{\sum_{N=1}^{\infty}2N \rho(N)}=\frac{1}{\bar{N}},
\end{equation}
where the factors 2 appear because each molecule has two possible bonding sites. In the ideal free-association (IFA) model\cite{Ciferri} the polymers form an ideal gas, and each bond lowers the energy by a discrete amount $\epsilon$.  This results in $\rho(N)\sim \rho(1)^N$, and gives 
\begin{equation}\label{eq:barN}
 \bar{N}=\frac{1}{2}+\frac{1}{2}\sqrt{1+4\rho v \exp(\beta G)},\\
\end{equation}
where $G\lesssim \epsilon$ is an effective free energy per bond and $v$ is a bonding volume. Both parameters are not known a priori, and are usually fitted to the average chain lengths, determined by e.g. circular dichroism measurements\cite{jonkheijm}. The WTPT includes the spatial extension of the molecules \cite{Wertheim1} neglected in the IFA-model, by calculating a reference hard sphere free energy and adding the attractive contribution of the pair potential. WTPT also assumes that no rings can be formed and that only one interaction per attractive site is possible. The average chain length predicted by the WTPT is formally equal to the IFA-result of Eq.~\ref{eq:barN} if
\begin{align}
\Delta &\equiv 4\pi \int g_{\text{rep}}(r) \left\langle \exp(-\beta V_{\text{att}}(r))-1  \right\rangle_{\omega_1,\omega_2} r^2dr \nonumber\\
&=\frac{v}{2}\exp(\beta G)\label{eq:Wertheim}
\end{align}
where $\Delta$ involves a single site-site interaction, and is related to the second virial coefficient\cite{sciortino}, $\langle.\rangle_{\omega_1,\omega_2}$ denotes an average over all orientations $\omega_1$ and $\omega_2$ of the two molecules, $g_{\text{rep}}(r)$ is the pair correlation function of the hard-sphere part of the potential, and $V_{\text{att}}(r)$ is the attractive part. At low densities and temperatures the IFA-model and the WTPT are equivalent, and in that case, for the square-well patchy particle of Ref.~\cite{sciortino}, $G$ is simply the well depth and $v$ can be calculated analytically. 

In Figure 2 we compare the aggregation fraction $\eta$ as a function of temperature $T$ for several values of the flexibility $w$ to the prediction of the WTPT with $g_{\text{rep}}(r)=1$, appropriate at low densities, and $\Delta$ from Eq.~\ref{eq:Wertheim} numerically calculated. The remarkable agreement with simulations shows that the WTPT also holds for smoothly varying potentials on a hard sphere. For comparison we also show a fit of the IFA-model, that deviates at higher temperatures due to the temperature dependence of the association energy. We find (not shown) that this deviation reduces with decreasing $w$, and is negligible for $w=0.1$. The polymerization temperature $T^*$, defined as the temperature where half of the molecules in the system has aggregated, i.e. where $\eta(T^*)=\frac{1}{2}$, increases with $w$. This rise in $T^*$  is due to an increase in available bonding volume $v$, or, equivalently, because a transition from an unbound to a bound state costs less entropy for a more flexible chain. For $w=0.4$ and $k_BT<0.058\epsilon$ the aggregation fraction suddenly jumps to $\eta\sim 1$. The chains have bundled, with a concomitant increase of the average chain length and a strong depletion of the gas density. Such a sudden increase of aggregation, not accounted for by polymerization theory, has recently been observed for the OPV-4 molecule\cite{jonkheijm}. Moreover, the transition is reminiscent of one of the assembly pathways suggested for the formation of amyloid fibrils \cite{Goldsbury}. 

\begin{figure}[t!] 
\begin{center} 
\includegraphics[width=\columnwidth,keepaspectratio]{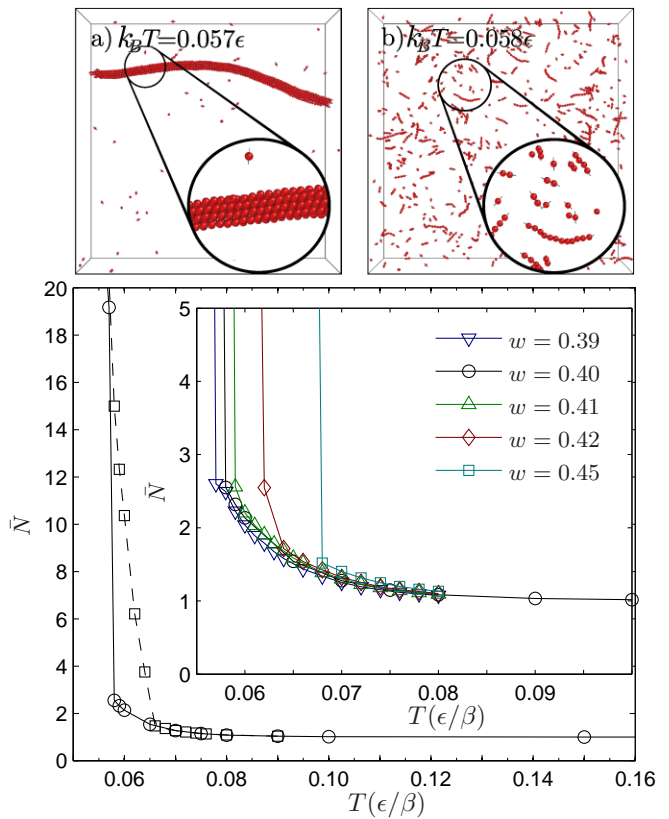}
\end{center} \caption{a) Top panels: representative configuration for $w=0.4$ at two very close temperatures. $M$, $\rho$ as in Fig.~2. a) $k_BT=0.057\epsilon$, where a solid bundle has nucleated. b) $k_BT=0.058\epsilon$ where the system is still a polymer gas. c) Average chain length $\bar{N}$ as a function of temperature for $w=0.4$. The solid line shows the jump of the chain length for $k_BT\sim 0.057\epsilon$ on cooling. The dashed line shows the hysteresis of the transition while raising the temperature from an equilibrated system initially at $k_BT=0.057\epsilon$. Inset: Same curve for several values of $w$. No hysteresis loop is shown. Notice that the bundling temperature increases with $w$. \label{fig:fig3}}\end{figure}

In Fig.~3 we examine the bundling transition in more detail. Figures 3a and 3b show the dramatic difference between a bundled and a polymer gas configuration, while the temperatures differ by only $k_B\Delta T=0.01\epsilon$. In the bundle the individual chains remain identifiable, with no bridging connections between chains. 
Visual inspection of configurations in the process of bundling, shows that when three chains come together, they remain bonded and suddenly grow in length, suggesting a nucleation mechanism. In Figure 3c the average chain length $\bar{N}$ is shown as a function of temperature for $w=0.4$, where the bundling transition is visible as a sharp jump. The dashed line represents systems heated up from a bundled configuration at $k_BT=0.057\epsilon$. This hysteresis together with the nucleation mechanism, is  evidence for a first order phase transition. We identify the actual bundling temperature $T_b$ at the high end of the hysteresis-loop, at $k_BT_b\approx0.065\epsilon$. Nevertheless, only at $k_BT=0.057\epsilon$ the critical nucleus is small enough to appear spontaneously during the duration of one simulation. In the inset of Fig.~3c we show $\bar{N}$ as a function of $T$ for several values of $w$. Although a stiffer chain loses less entropy upon bundling than a flexible one, $T_b$ increases with flexibility $w$, similar to $T^*$. The transition is thus not driven by entropy, but by the interaction energy of neighboring chains (Eq.~\ref{eq:patchLJgauss}) that, at large $\theta$, increases with $w$. Assuming that the patches in neighboring chains are perfectly aligned, the smallest angle $\theta_i$ between molecules in neighboring chains is of the order $\theta_i\sim\arctan(2^{-\frac{1}{6}})\approx 0.73\text{rad}$ or $42^\circ$. For  $w\lesssim 0.15$ this lateral interaction is negligible ($\theta_i/2w\approx 2.4$). Increasing $w$  also increases the available lateral bonding volume, decreasing the entropy loss upon bundling.

\begin{figure}[t!] 
\begin{center} 
\includegraphics[width=\columnwidth,keepaspectratio]{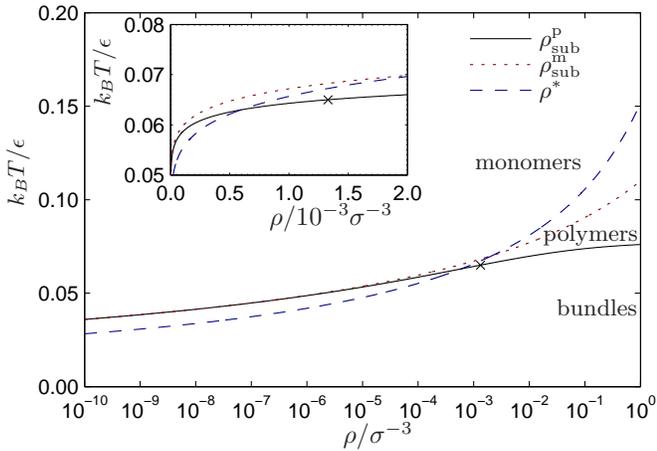}
\end{center} \caption{The gas solid coexistence line (sublimation line) with $G=0.8 \epsilon$, $Z_1=1$, $\epsilon_\text{solid}=-1.3$ for $w=0.4$, $v=3e-3\sigma^3$, and $\rho=0.0013\sigma^{-3}$ fitted to $k_BT^*=0.065\epsilon$ resulting in $k=2.1\cdot 10^3\epsilon\sigma^{-2}$. The solid line is calculated from Eq.~\ref{eq:coexistence1}, the dotted line from \ref{eq:coexistence2}, the dashed line from Eq.~\ref{eq:polline}. In the inset we show the same, but on a linear density scale.  \label{fig:fig4}}\end{figure}

We interpret the bundling as a sublimation transition, from a gas of polymers to a solid bundle. To derive an approximate equation of state, we equate the chemical potential $\mu_\text{sol}$ of a bundle to that of a polymer gas $\mu_\text{gas}$. Using the IFA-model we can derive 
\begin{equation}
 \mu_\text{gas}=\frac{\partial F}{\partial M}=k_B T \ln\left(\frac{v \rho}{Z_1}\right)  - 2\beta^{-1}\ln\bar{N}\label{eq:finalmu} 
\end{equation}
where the first term on the right hand side is the chemical potential of an ideal gas of monomers with internal partition function $Z_1$. For our rigid molecules, $Z_1=1$. As a first approximation, we model the bundle by an Einstein crystal\cite{FrenkelLadd} with
\begin{equation}
\mu_\text{sol}=\epsilon_\text{sol}+k_B T\left[ \ln\left(\frac{\Lambda^3}{\sigma^3}\right) +\frac{3}{2}\ln\left(\frac{ k\sigma^2}{ 2\pi k_B T}\right)\right]\label{eq:musolid1}
\end{equation}
where $\epsilon_\text{sol}$ is the well depth, including the association energy $G$ and the various chain-chain interactions of a molecule in the bundle. Furthermore,  $k$ is the spring constant of the Einstein crystal. We replace the de Broglie wavelength term with the bonding volume  $\Lambda^3=v$, as was done to derive the IFA-model. 
We measure $\epsilon_\text{sol}$ in a simulated bundle and fit $k$ to the sublimation transition of Fig.~3 ($\rho=3\cdot 10^3\sigma^{-3}$, $k_BT_b=0.065\epsilon$). Solving $\mu_\text{gas}=\mu_\text{sol}$ yields the sublimation density $\rho^\text{p}_\text{sub}(\beta)$
\begin{equation}
 \rho^\text{p}_\text{sub}(\beta)=\frac{\exp(-\beta\mu_\text{sol})Z_1}{v\left[Z_1\exp(\beta G)-\exp(-\beta \mu_\text{sol})\right]^2}.\label{eq:coexistence1}
\end{equation}
Equating the chemical potential of an ideal gas to $\mu_{\text{sol}}$ of Eq.~\ref{eq:musolid1} results  in the sublimation line for an ideal gas of monomers $\rho^\text{m}_\text{sub}(\beta)$ 
\begin{equation}
\rho^\text{m}_\text{sub}(\beta) =\left(\frac{\beta k}{2\pi}\right)^{\frac{3}{2}}\exp(\beta \epsilon_\text{solid}) \label{eq:coexistence2}
\end{equation}
It is useful to also define the density at the polymerization temperature $\rho^*=\rho(T^*)$, by combining Eqs.~\ref{eq:barNeta} and \ref{eq:barN}
\begin{equation}
\eta(T^*)=\frac{1}{2}\quad\Rightarrow\quad \rho^*=\left(2-\sqrt{2}\right)v^{-1}\exp(-\beta G).\label{eq:polline}
\end{equation}
Comparison of $\rho^\text{p}_\text{sub}$ to $\rho^*$ allows us to estimate whether the sublimation transition is dominated by the bundling of polymers (Eq.~\ref{eq:coexistence1}), or of monomers (Eq.~\ref{eq:coexistence2}).
 
In Figure 4 we compare the sublimation line of a polymer gas $\rho^\text{p}_\text{sub}$ to that of an ideal gas of monomers $\rho^\text{m}_\text{sub}$ in coexistence with the same Einstein crystal, and the polymerization line $\rho^*$.  At low densities $\rho^\text{p}_\text{sub}\sim \rho^\text{m}_\text{sub}$, because the transition from gas to solid occurs at a higher temperature than polymerization, i.e. $T_b>T^*$. The sublimation lines differ at higher densities, because $\mu_\text{gas}$ tends to the binding energy $G$, whereas the chemical potential of an ideal monomer gas tends to zero. At high densities polymerization occurs at higher temperatures than bundling, i.e. $T_b<T^*$. Sublimation requires $\epsilon_\text{sol}<-G$. As decreasing $w$ lowers $\epsilon_\text{sol}$, we expect a shift of the sublimation line $\rho_\text{sub}$ to lower temperatures, increasing the polymer-dominated regime. When the lateral interactions vanish, i.e. below $w=0.15$, the bundling transition will completely disappear.

In summary, we have presented a model based on patchy particles, that describes supramolecular polymerization and displays a first order phase transition to bundles, due to weak chain-chain interactions. The chain-to-bundle transition can be seen as a sublimation transition from a polymer gas to a solid bundle, for which we have been able to give a qualitative equation of state. We have related the occurrence of the phase transition to the flexibility of the supramolecular polymer. Our simulations show that bundling leads to a sudden increase of the average length of the aggregates, as experimentally observed in different polymerizing systems, from OPV-4 to amyloids.

We believe that the patchy particle is a flexible, powerful tool for efficient modeling of complex, self-assembling systems that can be adjusted at will by including more specific, e.g. chiral, chain-chain interactions. 
\acknowledgments
This work is a part of the research programme of
the ''Stichting voor Fundamenteel Onderzoek der Materie
(FOM),`` which is financially supported by the ``Nederlandse
Organisatie voor Wetenschappelijk Onderzoek (NWO).'' We thank F.~Sciortino and E.~Bianchi for useful discussions.
\bibliographystyle{achemso}

\end{document}